\documentclass[
,draft            
aas_macros]
{aipproc}

\layoutstyle{6x9}

\newcommand{\BATSE}{\textit{BATSE}}

\begin{document}

\title{A Magnetar Flare in the \BATSE~Catalog?}


\author{A. Crider}{
  address={Elon University, CB 2625, Elon, NC 27244}
}

\begin{abstract}
To identify extragalactic magnetar flares, we have searched for their periodic tails by generating Lomb periodograms of the emission following short bursts detected by the \textit{Burst and Transient Source Experiment} (\textit{BATSE}). 
Out of 358 short bursts examined, one has a significant tail periodicity ($T$ = 13.8 s,  ${P= 4 \times 10}^{-5}$).  The most probable host galaxy for this burst is ``The Fireworks Galaxy'' NGC 6946 (\mbox{$d = 5.9$ Mpc}).   At this distance,  the energy of the spike, ${(2.7 \pm 0.3) \times 10}^{44}$ ergs, is akin to those of the galactic magnetar giant flares, as are its duration ($\sim 0.4$ s) and temperature ($250 \pm 60$ keV).  For the tail emission, however, our estimated temperature of $60 \pm 5$ keV is harder and the energy release of ${(4.3 \pm 0.8) \times 10}^{45}$ ergs is larger than those of the galactic magnetar flares.  Regardless of the host, such a large ratio of tail-to-spike energy would imply that magnetar flare tails might be detectable out to further distances than previously thought. 
\end{abstract}

\maketitle


\section{Introduction}

Three times within the past thirty years, short (0.2-0.35 s), intense gamma-ray flares followed by softer ($kT \sim 25$ keV), periodic ($T=5-8$ s) tails have erupted from the sources of the much fainter soft-gamma repeaters (SGR).  The magnetar model proposed by Duncan and Thompson \citep{1992ApJ...392L...9D} has had considerable success in explaining SGR and the occasional giant flares. 
Assuming that our galaxy is not unique, then magnetar flares also occur in nearby galaxies.  These would likely be labeled as short GRB, since the characteristic pulsating tail would near or below the background and the spectrum of the initial spike is similar to that of classical GRB \citep{1996ApJ...460..964F}.  To identify them, one can exploit three distinguishing attributes: (a) their locations relative to nearby galaxies, (b) their spectral temperatures, and (c) their faint oscillating tails.  In this study, we searched for faint oscillations in the emission following short GRB.

\section{Procedures}

The current \BATSE~catalog contains 2702 gamma-ray bursts; 2041 have calculated $T_{90}$ durations.   For the 358 short bursts (\mbox{$T_{90} < 1.0$ s}), we extracted the 30-50 keV, 64-ms lightcurve (as the galactic magnetar flare tails were soft) and refit polynomial backgrounds using data from 100 s before to 200 s after the trigger.   We then generated a Lomb periodogram \citep{Numerical-Recipes} for the 100-s intervals that immediately followed each short GRB.  
Of the 358 bursts analyzed, only one had a significance $P < 1/358$. The periodogram plotted in Figure 1 for GRB 970110 (\BATSE~\#5770) reveals a highly significant (\mbox{Lomb power=17.8}, \mbox{$P = {3 \times 10}^{-5}$}) peak periodicity of 13.8 s.  We found the signal separately in both of the two \BATSE~Large Area Detectors (LADs) facing the event.    Examining the 50-100 keV channel independently reveals the same strong periodicity (Lomb power=16.2, $P = {1 \times 10}^{-4}$).    No significant signal was found in the upper two channels (>100 keV), as might be expected for a  soft magnetar flare tail.  Figure 2 shows the countrate rebinned from 64-ms to 2-s bins and overlayed with a 13.8-s sinusoid to illustrate the periodicity found originally by the Lomb periodogram.

\begin{figure}
\includegraphics[height=.4\textheight]{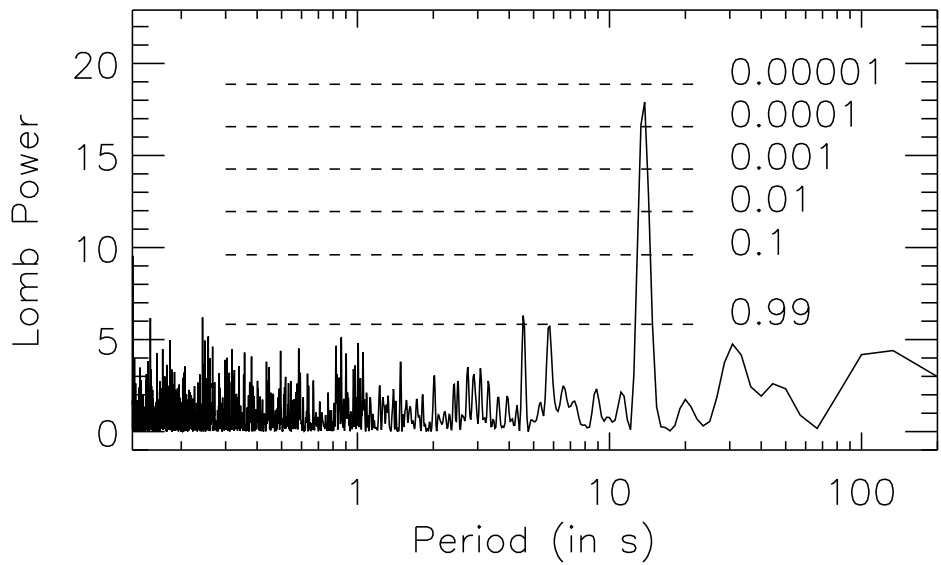}
\caption{Lomb periodogram of the 100-s interval following GRB 970110.  The dominant periodicity of 13.8 s has a Lomb power of 17.8.  Significance levels are plotted as horizontal lines.  The low chance probability of such an intense signal (\mbox{$P = {3 \times 10}^{-5}$}) implies that this feature is real.}
\end{figure}

To test if this periodic signal was from a source unrelated to the burst, we examined periodograms of the pre-burst (-100-to-0 s) and subsequent (100-to-200 s) regions and found no significant periodicity, with maximum significances of $P=0.55$ and 0.45 respectively.  While the 1.024-s binning of the pre-burst emission limits the sensitivity of the Lomb periodogram to some extent, the period of interest (0-to-100 s) retains a marginally significant spike ($P=0.01$) when resampled to this resolution.  Curiously, there may be pulses in phase with the tail during the 50 s before the trigger, as can be seen in Figure 2.  However, the lack of a sustained periodic signal before or 100 s after the trigger suggests that the pulsations are indeed a transient phenomena temporally coincident with the spike.  Additionally, we found no hint of a 13.8-s periodicity in the DISCLA data of the other six LADs, suggesting the source is indeed from the direction of the burst.  The Australia Telescope National Facility (ATNF) Pulsar Catalog \citep{2005AJ....129.1993M} that includes the known gamma-ray pulsars and the anomalous X-ray pulsars contains no sources with a periodicity $> 2$ s inside the 99.7\% confidence location of GRB 970110.     While Cyg X-1, in its ''soft/low'' state on 1997 January 10, is just outside of the 99.7\% \BATSE~confidence circle, there are no reports of it having a periodicity close to 13.8 s.

\begin{figure}
\includegraphics[height=.35\textheight]{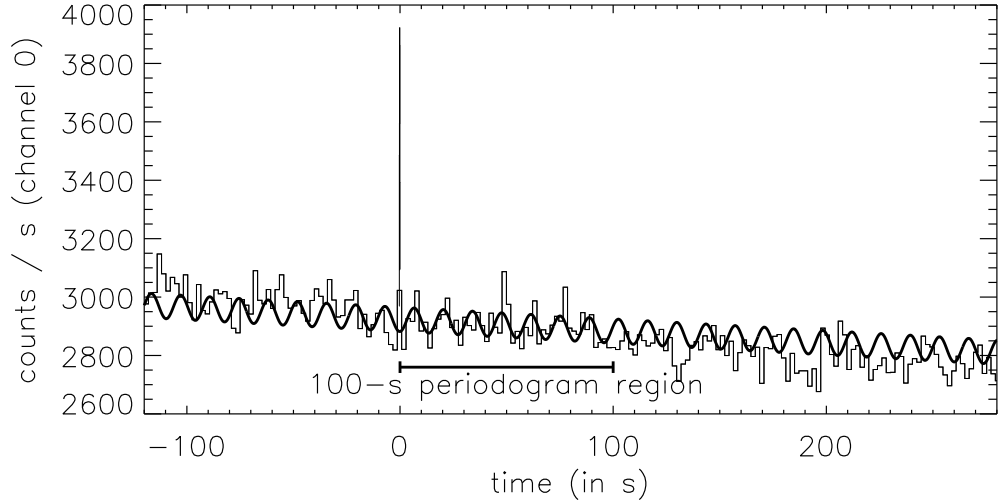}
\caption{The \BATSE~30-50 keV count rate around GRB 970110 at 2-s resolution.   The best fit sinusoid with a 13.8-s period is plotted for reference.   The burst itself is plotted at 64-ms resolution.}
\end{figure}

The duration of the spike ($\sim 0.4$ s) is very similar to those of the magnetar flares.  
Using NASA-MSFC's \textit{rmfit} analysis software and \BATSE~DISCSC data, we also found a similar spike spectrum.  An optically-thin thermal bremsstrahlung (OTTB) spectrum gives an acceptable fit ($\chi^{2} = 1.32, \nu=2, P = 0.52$) with a temperature ${kT}_{\rm{OTTB}} = 250\pm60$ keV.   A blackbody spectrum fits more poorly ($\chi^{2}  = 8.2, \nu=2, P = 0.017$) with ${kT}_{\rm{BB}} = 30$ keV.  Our result is consistent with the peak of the 1979 March 5 event (${kT}_{\rm{OTTB}} =246$ keV; \citep{1996ApJ...460..964F}), but is softer than the the 2004 December 27 flare (${kT}_{\rm{BB}} = 175\pm25$ keV; \citep{2005Natur.434.1098H}).  Using $rmfit$, we found the fluence of the spike to be ${(6.5 \pm 0.5) \times 10}^{-8}~\rm{erg~{cm}^{-2}}$ in \BATSE's 25-2000 keV window. 

Determining the tail's spectrum and fluence required calibrating the measured Lomb power and the total detector counts in each channel.
We created several synthetic magnetar tails using the \textit{Swift} BAT light curve for the 2004 December 27 magnetar flare ($t= 205$ to 505 s).
These were scaled to represent what would be seen in each \BATSE~channel, with total counts ranging from 200 to 400,000.   
A periodogram for each allowed us to estimate the functional relationship (roughly quadratic in the region of interest) between the integrated detector counts in a channel and the Lomb power .  We found our Lomb powers of 17.7 in channel 0 and 16.2 in channel 1 corresponded to 6000 counts and 5100 counts, respectively.  We next convolved OTTB spectrum ($kT_{OTTB}$ spanning 5 to 70 keV) with the detector response matrix for  \BATSE~trigger \#5770 to determine the number of counts expected in the four channels.    From this, we estimated that the temperature of the tail is ${kT}_{\rm{OTTB}} = 60\pm5$ keV, notably harder than other magnetar flares.  With this spectrum, and the Lomb power for channel 0, we estimate a tail energy fluence of ${(7.5 \pm 1.5) \times 10}^{-7}~\rm{erg~cm}^{-2}$.  The fraction of the total energy in the tail emission (94\%) is much larger than the recent 2004 December 27 event (0.3\%), but is comparable to the 1979 March 5 event (75\%).

\section{Discussion}

The \BATSE~localization of GRB 970110 includes few nearby galaxies.  
The dwarf spheroidal galaxy Draco ($d$ = 0.08 Mpc) and the blue compact dwarf galaxy NGC~6789 ($d$ = 3.6 Mpc) both fall just inside of the 95.4\% confidence circle, but have low \textit{a priori} probabilities of producing a magnetar flare based on their low star formation rates.  Instead, we find a Bayesian probability of 87\% that of the galaxies within 10 Mpc,  the ``Fireworks Galaxy'' NGC 6946 ($d$ = 5.9 Mpc) is the host.  While just outside of the 95.4\% confidence circle, its very high star formation rate of 3.12 $M_{\odot}~{\rm{yr}}^{-1}$ \citep{2005A&A...434..935K} makes it a likely source.
In fact, we estimate a 38\% \textit{a priori} probability that a magnetar flare from \mbox{NGC 6946} exists in the \BATSE~catalog. Assuming isotropic emission, the energy fluence in the spike corresponds to \mbox{${(2.7 \pm 0.3) \times 10}^{44}$ erg}, comparable to 1979 March 5 flare, which had a spike energy of \mbox{${1.2 \times 10}^{44}$ erg} \citep{1979Natur.282..587M}.   Its tail energy of ${(4.3 \pm 0.8) \times 10}^{45}$ ergs is larger than those of the three galactic magnetar flares, but only $\sim 10\times$ more than that of the 1979~March 5~event.   The requisite dipole magnetic field strength of the magnetar that would constrain such a fireball would be~
\begin{math}
{B_{\star}}~>~{1.4 \times 10}^{15}
~({ {E}_{\rm{tail}} }/{ {4.3 \times 10}^{45}~{\rm{erg}} } )^{1/2}
~({\Delta R}/{10~{\rm{km}}})^{-3/2}
~{({1/2 +\Delta R / 2R_{\star}})}^{3}
\rm{G},
\end{math}
where $R_{\star}$ is the stellar radius and $\Delta R$ is the outer radius of the magnetic loop confining the plasma  \citep{1995MNRAS.275..255T}.
If instead GRB~970110 is from the Draco dwarf galaxy, then its energetics are more similar to the event that occurred two days after the 1997~August~27 giant flare \citep{2001ApJ...558..237I}.  In either case, the relatively large fraction of energy in the tail of this event suggests that the range to which \textit{Swift} might detect similar periodicities should be extended.  \citet{2005Natur.434.1098H} calculated that magnetar periods might be measured by \textit{Swift} out to a distance of $\sim2-8.5$ Mpc based on the fluence observed for the 2004~December~27 event.  Approximately 15\% of the tail energy in the 1997~January~10 event (${6.5 \times 10}^{44}$ ergs) was released in the \textit{Swift} XRT band (0.3-100 keV), suggesting that if this event is from NGC 6946, the Swift detection range for tails can be extended to $\sim5-20$ Mpc.   

\begin{theacknowledgments}
The author thanks Michael Briggs, Dana Hurley Crider, and his \textit{PHY 251} students for helpful comments.  This work was supported by a grant from Elon University.
\end{theacknowledgments}



\bibliographystyle{aipproc}   

\bibliography{ads}

\IfFileExists{\jobname.bbl}{}
 {\typeout{}
  \typeout{******************************************}
  \typeout{** Please run "bibtex \jobname" to optain}
  \typeout{** the bibliography and then re-run LaTeX}
  \typeout{** twice to fix the references!}
  \typeout{******************************************}
  \typeout{}
 }

\end{document}